\documentstyle[preprint,aps]{revtex}
\tighten
\begin{document}
\widetext
\input{epsf.sty}
\input epsf
\bigskip
\vspace{0.2in}
\thispagestyle{empty}
\begin{flushright}
NYU-TH/99/3/02\\
\today 
\end{flushright}

\vspace{0.2in}

\begin{center}
\bigskip\bigskip
{\large \bf Non-conservation of Global Charges 
in the Brane Universe
\\ \vspace{0.1in} 
and Baryogenesis}
\vspace{0.3in}      

{Gia Dvali\footnote{Also, International Center for 
Theoretical Physics, 34100 Trieste, Italy},~~~Gregory Gabadadze}
\vspace{0.2in}

{\baselineskip=14pt \it 
Department of Physics, New York University, New York, NY 10003, USA} \\
{\it email: dvali, gabadadze@physics.nyu.edu } \\
\vspace{0.2in}
\end{center}

\vspace{0.9cm}
\begin{center}
{\bf Abstract}
\end{center} 
\vspace{0.3in}

We argue that global charges, such as baryon or lepton number,
are not conserved in theories
with the Standard Model fields localized on the  brane which 
propagates in higher-dimensional space-time.
The global-charge non-conservation is due to quantum fluctuations 
of the brane surface. 
These fluctuations create ``baby branes'' that can capture
some global charges and carry them away into the bulk of 
higher-dimensional space. Such  processes are
exponentially suppressed at low-energies, but can be 
significant at high enough temperatures or energies.
These  effects  can  lead to a new, intrinsically high-dimensional mechanism
of baryogenesis. Baryon asymmetry might be produced due either to
``evaporation'' into the baby branes, or creation of the baryon number 
excess in collisions of two Brane Universes.
As an example  we discuss a  possible  cosmological
scenario within the recently proposed ``Brane Inflation'' framework.
Inflation is driven by displaced branes which
slowly fall on top of each other.
When the branes collide inflation stops and the Brane Universe
reheats. During this non-equilibrium collision 
baryon number can be  transported from one brane to another one.
This results in  the baryon number excess in  our Universe 
which exactly equals to the hidden ``baryon  number'' deficit 
in the other Brane Universe.

\newpage
\noindent 
{\bf 1. Introduction} 
\vspace{0.2in}

The only candidate theory for consistent quantum description of 
all the interactions in Nature can be formulated in space-time with
dimensionality greater than four. One possible scenario,
of how our four-dimensional world emerges in this picture, is
based on the assumption that all the observed particles are localized
on  a 3-dimensional brane 
that propagates in higher-dimensional space-time \cite {RubakovS}.
The absence of supersymmetry in the observable world 
can be related to a non-BPS nature of the brane \cite {DvaliShifman1}.
Within  the field theory context the simplest localization mechanism
for fermions is due to the index theorem in the
solitonic background \cite {JackiwRabbi}  which was first used in Ref. 
\cite {RubakovS} for a five-dimensional Brane Universe scenario.

As shown in \cite {DvaliShifman}, 
the localization of 
spin-1 gauge-fields in the field theory  context is less straightforward. 
The mechanism requires the gauge group to be confining 
away from the brane \cite {DvaliShifman}. 
This has an  important model-independent consequence
for the gauge-charge conservation on the brane, which is nothing but 
gauge flux conservation in the confining medium. 
The similar argument applied to spin-2 particles
within the field theory context  
shows why localization of gravitons is hard to achieve. 
Thus,  gravitons,   unlike other particles 
must live in the brane as well as in the bulk of extra 
compactified dimensions.

One of the motivations for the  Brane Universe scenario is
that it allows to lower the fundamental scale of quantum
gravity all the way down to TeV or so, thus, providing a novel
view on  the Hierarchy problem \cite {ADD,AntoniadisCo}. 
The observed weakness of gravity
at large distances  is because 
gravitational fluxes spread into the $N$ extra dimensions. 

The relation between the observed Planck scale, $M_{\rm P}
\simeq 10^{ 19}~{\rm GeV}$,  and the so-called 
fundamental  Planck scale $M$ is then given by \cite {ADD}:
\begin{equation}
M_{\rm P}^2 = M^{N + 2}V_N \label{planckscale}
\end{equation}
where $V_N \sim R^N~~{\rm with}~~N\ge2 $, is the volume of extra
compactified  spatial dimensions.
The size of compactified radii in this picture can be in a sub-millimeter
range  without conflicting with any present astrophysical or laboratory
constraints \cite {ADD1}.  

Perhaps, the best motivated framework for
the Brane Universe scenario can be found  within 
superstrings  and ${\rm M}$ theory.  
In fact, D-branes \cite {Dbranes} 
provide natural candidates for objects on which 
the observed particles could live. 
Thus, the Standard Model particles can be identified
with massless modes of open strings which are ``stuck'' to  a
set of overlapping D-branes. 
Gravity, on the other hand, is given by the closed string sector
of the theory.  
It was shown in Refs. \cite {AntoniadisCo,Antoniadis}
that the TeV scale quantum gravity models can be 
embedded in superstring theories
(this approach was named ``Brane World'' in Ref. \cite {ZuraTye})
\footnote{The idea 
to lower the the Planck scale to the unification  scale originates in  
\cite {Witten}; An attempt to lower the string scale to the electroweak
scale  was discussed in Ref. \cite {Lykken}.}. 
Various phenomenological aspects of these models are  
studied in the literature (for an incomplete list 
of references see, e.g., \cite {Rattazzi,Peskin,jgro,Dines,Cosmology1}).  

In the present work we argue that global charges are 
not conserved in the Brane Universe. 
The  non-conservation of 
global charges is due to quantum 
fluctuations of the brane on which the Standard Model lives. 
These fluctuations can produce baby branes which can capture
global charges and carry them away from the brane. 
At high enough temperatures  or energies comparable  with 
the brane tension the process  of baby brane creation 
becomes significant. This leads to the global-charge  transport 
from our brane. The corresponding process will look
as non-conservation of global charges for a four-dimensional observer
living on the brane\footnote{These processes somewhat resemble
the loss of quantum coherence in quantum gravity \cite {W1,W2,W3}.}. 
These  non-conservation mechanisms are  
significant in the  cosmological context and can 
lead to new sources of baryogenesis.
We discuss a possible  cosmological
scenario based on recently proposed Brane Inflation mechanism 
\cite {DvaliTye}. This scenario 
results in the baryon number access in  our four-dimensional 
Brane Universe.  

The terminology we adopt in this work  is  as follows.
The three-dimensional brane
on  which we live will be referred as  the mother brane. 
The $(3+1)$-dimensional Universe accommodated on the mother brane 
will be called the Brane Universe. 
The Brane Universe is able to
communicate gravitationally and/or via some other 
bulk fields with other Brane Universes which also
propagate in the original higher-dimensional (presumably
ten or eleven-dimensional) space-time.   
This feature is going to be decisive for what we discuss below.
In section 2 we discuss a number of mechanisms of non-conservation
of global charges in the Brane Universe.   
We  show (subsection 2.1) that at high enough temperatures, or 
in scatterings at energies of order TeV or so,  
the Brane Universe could  produce baby branes which would carry the
Standard Model global quantum numbers
off the Brane Universe. 
For an observer living in the Brane Universe
these processes will look as non-conservation of 
global quantum numbers,
such as non-conservation of baryonic  or leptonic  charges.
In subsection 2.2  a field-theoretic  scenario  
of the global charge non-conservation 
in the hot Brane Universe is proposed. 
The Brane Universe, at some finite temperature
produces confining bubbles with net global charges 
and liberates  them into the bulk of higher-dimensional space.
This scenario is complimentary to the one discussed in subsection 2.1.
In section 3 we discuss applicability of these mechanisms
to the baryon asymmetry problem.   
We argue that within the cosmological scenario
proposed in \cite {DvaliTye} these non-conservation
processes should lead to the baryon 
asymmetry in the Brane Universe. 
Discussions and conclusions are 
presented in section 4. 
\vspace{0.2in} \\
{\bf 2. Non-conservation of Global Charges in the Brane Universe}
\vspace{0.2in}

In this section we  propose mechanisms for 
non-conservation of global charges in the Brane Universe. 
The objective is to demonstrate that there are intrinsically
high-dimensional 
phenomena which lead to baryon and lepton number non-conservation on
our brane.
\vspace{0.1in} \\
{\it  2.1. Baby Branes and  Non-conservation of Global Charges}
\vspace{0.1in}

Let us start with the case when our 
$(3+1)$-dimensional Brane Universe is embedded in higher 
dimensional space-time. We will think of the brane as being a dynamical
object with the tension $\sigma  \sim M^4$.
This brane can fluctuate.  The fluctuations are stronger 
at high temperature. In fact, there 
is a probability for the Brane  Universe to wiggle strongly and 
create a baby brane (see Fig. 1).
At high enough temperature  the baby brane will be able to 
pull off the mother brane and propagate 
into the bulk of higher-dimensional space.
This probability is non-vanishing if the temperature of the 
Brane Universe is nonzero (the same process could also be
seen at high enough energies).  
The rate of this process is exponentially suppressed and can be 
estimated as \cite {Linde}:
\begin{eqnarray}
{\Gamma \over {\rm Volume}}~ \propto ~{\rm exp} \Big
(-{E_b\over T}\Big )~,  
\label{rate}
\end{eqnarray}
where $E_b$ stands for the surface energy of the baby brane, which 
can be determined  via  its surface area  $A$ and tension $\sigma$,
$E_b=A\sigma$. $T$ stands for temperature of the 
Brane Universe. Thus, for high temperatures 
of order TeV or so, creation of baby branes should be an 
appreciable  effect, while it should  drop off rapidly as 
the brane cools down.

Once the baby brane  is formed, it  can capture some particles
which happen to be nearby  
and carry them away from the mother brane.
What will happen if the captured particles carry 
a net global charge, let us say  baryon or lepton number?
To clarify the issue,  
let us consider the case 
when the captured state  is a gauge singlet 
combination of $u_R, d_R~{\rm and}~d_R$ quarks (it might be 
any other combination  
of the Standard Model states  which carries  
a non-zero baryon or lepton number but 
has  {\it strictly zero} gauge charge). 
For a four-dimensional observer living on our  brane
this process will look as follows:
\begin{equation}
u_R + d_R + d_R \rightarrow {\rm NOTHING},
\end{equation}
where "NOTHING" stands for  the baby brane 
which got separated from the mother brane and carries away  the
corresponding global charge.  
Since this object can  gravitate,  
it will look for a 
four-dimensional observer as a piece of dark or hidden matter. 
Thus, the baby brane will carry its own baryon
number   $B_{\rm baby}$.  
The value of $B_{\rm baby}$ will exactly equal to 
the baryonic  charge that is lost on the mother brane. 
If there are no bulk particles which can carry 
well defined baryon number, 
an observer on the mother brane 
will not be able to measure  $B_{\rm baby}$.
Thus,  the process will look as disappearance of baryonic  charge
$\Delta B$ on the mother brane 
and appearance of the same charge $B_{\rm
baby} = -\Delta B$ on the baby brane\footnote{If there are 
some bulk particles that can carry baryon  number, the issue 
becomes  more subtle. In this case  $B_{\rm baby}$ can be measured
via the exchange of these bulk modes. If these modes are heavy
the induced effective interaction between the baryons localized on 
different branes will be exponentially suppressed by the Yukawa
factor ${\rm exp} (-r m)/r^{N-2}$. In this section we
will assume that there are no such particles in the bulk.}. 

Let us try to understand this effect from the point of view of the
effective four-dimensional  field theory which is seen at distances 
larger than the size of extra dimensions. 
Let $\Psi_M$ be a wave-function describing a state of
some group of particles on our brane which carries a net baryonic charge. 
Likewise,  we can define $\Psi_B$ to be a wave-function 
of a set of  particles on the baby brane. 
Both $\Psi_M$ and $\Psi_B$ are sharply
localized functions in the bulk. 
The overlap between them is exponentially suppressed
as the separation of the branes, $r$,  in extra dimensions increases.
Baryon number
conservation in the theory is a result of  the symmetry under the 
global phase rotations $\Psi_M \rightarrow {\rm e}^{iQ_M\alpha} \Psi_M$,
and $\Psi_B \rightarrow {\rm e}^{iQ_B\beta} \Psi_B$, 
where $Q_M$ and $Q_B$ denote the charges for the 
corresponding wave-functions. 
When the branes are separated the wave-functions $\Psi_M$ and $\Psi_B$
get decoupled. 
\begin{figure}[h]
\epsfbox{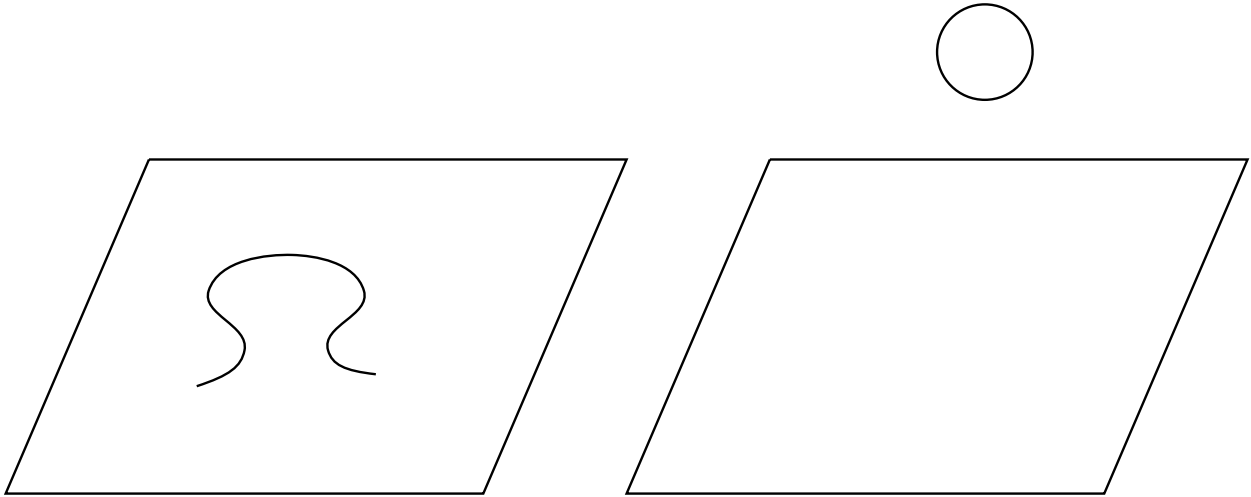}
\vspace{0.3in} 
\caption{{\small Creation of baby branes }} 
\label{fig1}
\end{figure} 
As a result, 
there are two independent $U(1)$ symmetries available: 
$U(1)_M$ and $ U(1)_B$.
However, when branes are close to each other, the functions  
$\Psi_M$ and $\Psi_B$ overlap. As a result, there is 
only one unbroken  combination of $U(1)_M$ and $U(1)_B$, call it
$U(1)_{\rm baryon}$,  
which is defining baryon number  of the whole  
system of the overlapping branes. 
In terms of the effective field theory language, the
effective low-energy Lagrangian 
contains $U(1)_M\otimes U(1)_B$ violating
term  whose strength depends on the distance between 
the branes.  
This term  drops
exponentially fast as the separation between the branes increases:
\begin{equation}
({\bar \Psi}_M) ^{Q_B}~ (\Psi_B)^{Q_M}~ {\rm e}^{-rM}.
\label{baryon}
\end{equation}
What is important here  is that the interaction should necessarily respect
the $U(1)_{\rm baryon}$ symmetry. 
As we mentioned above, for small $r$ branes are
interacting. Thus, 
there is only one baryon charge. This charge can be exchanged
among the states of  $\Psi_M$ and $\Psi_B$. 
Suppose $\Delta Q$ denotes the amount of charge which 
is being transferred from the one set to another one. 
Once the branes are  
separated ($r \rightarrow \infty$) the overlap term disappears.
Thus, there are two separately conserved charges corresponding to
$U(1)_M$ and $U(1)_B$ respectively.  However, 
only the charge  $Q_M$ will be seen in the mother brane 
and, thus, interpreted as the baryon number of our brane. 
Summarizing,  the
charge transport from $\Psi_M$ to $\Psi_B$ will look as
disappearance of the 
$\Delta Q$ amount of the baryon charge on our brane and 
as appearance of exactly the same amount of the baryon 
charge on the baby brane. 

Evidently, in each individual process $\Delta Q$ can take
either sign and, if the system is in equilibrium, the net baryonic charge
left on the brane will average to zero. 
However,  it might be possible 
to generate a net baryon asymmetry on the mother brane 
if the system was {\it out of}
equilibrium for some time during its evolution (it also requires
C and CP violation \cite {Sakharov}, see discussions below).
In section 3 we address this issue and  
propose possible scenarios  of how the baryon asymmetry
could be generated in the Brane Universe.

Before we turn to the next section, let us discuss the fate of  
local charges in the Brane Universe. 
Seemingly, the same non-conservation process might be happening with  
the gauge charges, such as  electric charge for instance.
However,  this cannot  be true \cite{DvaliShifman,ADD}. 
Indeed, consider the case
when the local charge is attached to 
the strongly fluctuating region of the mother brane 
which is about to be pulled  off (Fig.2).  The local charge, 
due to the  corresponding flux conservation,   
would necessarily create a flux tube originating at the location of 
this charge and ending on the mother brane (see Fig. 2).  
At high enough energies, or temperatures likewise,  the flux tube can  
break apart and the baby brane will eventually be liberated into  the 
bulk of higher-dimensional space. 
However, the liberated baby brane will necessarily be neutral
with respect to the local charge under consideration. 
Indeed, the process of breaking of the flux tube goes 
through creation of a charge-anti-charge pair in the tube.
Once this pair is created, 
the anti-charge will get  attracted by the original charge siting
on the baby brane. Thus, the flux tube will break apart in such a 
way that the anti-charge from the pair will be attached to the 
baby brane and the charge of the pair will be attached to 
the mother brane. Hence, the final configuration
of the liberated baby brane will be electrically neutral and the local 
charge will be conserved on the mother brane.
Another way of saying this is to recall that all the 
Standard Model gauge interactions should be in a confining 
phase in the bulk space-time \cite {DvaliShifman}.   

\begin{figure}[h]
\epsfbox{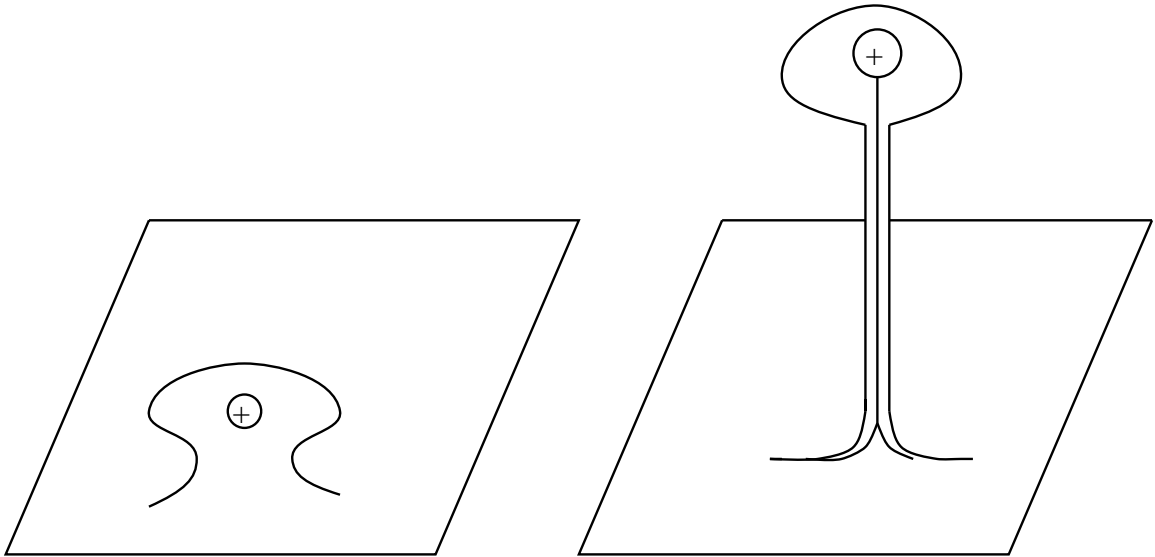}
\vspace{0.3in}
\caption{{\small Flux tube holding the baby brane with a local charge}} 
\label{fig2}
\end{figure}   

Summarizing the discussions in this section we conclude that
the process of baby brane creation should  
lead to non-conservation of
global charges (such as baryon or lepton number) in the Brane Universe.
Moreover, this process
will necessarily respect all the local charge  conservation laws.
\vspace{0.1in} \\
{\it  2.2.  Bulk ``Glueballs'' and ``Hadrons''} 
\vspace{0.1in}

It is important to emphasize that the existence of generic bulk
particles, which may carry baryon  number, cannot lead to the baryonic  
charge non-conservation in the four-dimensional effective field theory.
The reason why this was possible for the baby branes is that
branes are sharply localized coherent states, wave packets in some 
sense. Because of this property the baryonic charge localized on such 
an object is impossible to probe from a distant brane, and, thus,
is effectively lost.  
On the other hand, the wave-functions of the generic bulk particles,
such as Kaluza-Klein gravitons for instance, are spread over
the whole bulk and their baryonic charge 
can constantly be measured from our brane
at any energies. 

Another examples of such sharply localized objects  can be bulk
``glueballs'' or ``hadrons''. These are the states that appear in the bulk
due to the particular mechanism of 
localization of the gauge fields on the brane. 
As it was shown in \cite {DvaliShifman}, the field theory mechanism 
for localization of the massless gauge-fields on the brane implies that
corresponding gauge group is in a  confining phase in the bulk. 
Thus,  a pair of test charges places in the bulk should be connected by a 
flux tube with the  tension proportional to
$\Lambda^2$, where $\Lambda$ is a scale of the 
confining theory in the bulk. 
The inverse confinement scale,
$\Lambda^{-1}$, sets the localization width for  
the observed gauge fields. For 
phenomenological reasons $\Lambda$ should be greater than ${\rm TeV}$. 

Notice, that the gauge group  in the bulk can be bigger than the
Standard Model group. A photon, in this case,
if being emitted into  the bulk, becomes a gauge boson of the 
bigger  confining theory. Thus, the photon  can only escape the mother
brane in the form of a heavy bound state, a sort of bulk ``glueball''. 
The similar consideration
applies to fermion states. If the gauge group 
in the bulk were not confining, these fermions would have escaped 
the mother brane at energies bigger than the localization width. 
However, since the bulk is confining,  
such states  can only escape  
within the  corresponding ``colorless'' composite objects, bulk ``Hadons''.
Since the  bulk ``Hadrons'' might  carry off some net global charges, 
they can also lead to non-conservation of the global charges on the brane
\footnote{In all the discussions above and in subsection 2.3 
we assume for simplicity that the localization width for 
fermions approximately equals to $\Lambda$.}.  
\vspace{0.1in} \\
{\it 2.3. Confining Bubbles and Non-conservation
of Global Charges} 
\vspace{0.1in}

In this section we study  a field-theoretic 
model of non-conservation of global charges which is based on
confining properties of the bulk gauge group.  
As we mentioned before, the group 
is restored and confining outside of the 
brane \cite {DvaliShifman}.
For instance, the electroweak 
gauge group $SU(2)_L$ which is broken
in our brane, should be restored (or be a part of a bigger unbroken group)
and confining outside of the brane (see Fig. 3). 
The same applies to color $SU(3)_c$ and hypercharge 
$U(1)_{\rm Y}$ symmetries.
$SU(3)_c$ should either be 
a subgroup of bigger confining bulk gauge group, or
be the same bulk gauge group with the confinement scale 
greater than TeV. Likewise, $U(1)_Y$ should be 
a part of a bigger group that is confining in the bulk. 
For simplicity of arguments 
we will be assuming that the gauge group within  the 
brane is broken $SU(2)_L$ and outside of the brane it is 
confining  $SU(2)_L$ 
(the generalization to the other groups and interactions  
is straightforward).
Let us suppose that within the Brane Universe the confining phase which is
realized outside of the mother brane is seen as a local false vacuum
state. Then,  
in our four-dimensional world there is  a finite probability to 
create a bubble ( a sort of 
``hole'') with a confining phase inside. 
If some ``colorless'' states with nonzero global 
charges are captured inside the  bubble, 
they will be able to ``leak'' into the bulk. These effects are  
complimentary (but more model-dependent) to those discussed 
in the previous subsections. 
Let us study the bubble creation processes more carefully.   

The probability to create a bubble per unit volume 
per unit time in our world with the confinement phase inside of 
the bubble is given by \cite {VKO}:
\begin{eqnarray}
{P \over {\rm Volume}}~ \propto ~{\rm exp}\Big (-a~{\sigma^4 \over
(F(T)-{\cal E})^3 }\Big )~,~~{\rm when}~~{\bar \Lambda}^4>F(T)>{ \cal E}.  
\label{bubblerate} 
\end{eqnarray}
Here, $F(T)$ denotes the free energy of the system as a function of
temperature of the mother brane $T$, 
${\bar \Lambda}^4$ denotes  the depth of the scalar potential of the 
broken $SU(2)$ theory, ${\cal E}$ is the 
difference between the energy densities 
of the confining and the Higgs phases,  and 
$a$ stands for some positive constant of order 10-100.
As $T$ is close to $\Lambda\sim ~{\rm few~TeV}$
this probability becomes significant. The theory inside of the 
bubble is in a confinement phase. Thus, bound states of particles 
which might form  
within the bubble are to  be $SU(2)$  singlets\footnote{ 
If all the Standard Model interactions  are considered, 
these states are supposed to be 
``color singlets'' with respect to the whole Standard Model gauge group
or w.r.t. the corresponding GUT, if the unification is assumed.}.
These singlet states will be able to 
propagate  out of the Brane Universe.
The most dramatic signature of this propagation is that they 
will be able to carry
global quantum numbers off our Brane Universe. For instance, 
consider a single left-handed neutrino. This particle 
transforms  in the fundamental dublet of
$SU(2)_L$.
Thus, it carries a ``weak color'' charge and cannot escape the brane. 
However, in accordance with
't Hooft's correspondence principle \cite {tHooft}, the neutrino  of 
the theory with a broken $SU(2)_L$ can be thought of  as a 
``weak colorless'' state, or as a bound state of confining 
$SU(2)_L$. Indeed, in the confinement picture, the left-handed neutrino can
be presented as follows \cite {tHooft}:
\begin{eqnarray}
\nu_L~ {\rm in~ Higgs~phase}~<=>~{\bar H}^i L_i~{\rm in~ Confinement
~phase}~.
\label{neutrino}
\end{eqnarray}
 Here, $H$ stands for the Standard Model Higgs dublet,
$H_i^T=(\phi^+, \phi^0)$ and $L$ stands for
the left-handed dublet of a neutrino and electron, 
$L_i^T=(\nu_L, e_L)$. It is straightforward to see that the 
``weak colorless'' bound state ${\bar H}^i L_i $
reduces  to an ordinary left-handed neutrino once
the Higgs field is given a non-zero  vacuum expectation value (VEV).
Indeed, in the unitary gauge 
$H^T= {1\over \sqrt {2} } \Big ( v+h,~0 \Big )$, 
where $v$ denotes the Higgs VEV and $h$ stands for Higgs fluctuations
about this VEV. Substituting this expression into
the right hand side of Eq. (\ref {neutrino}) one finds,
$ {\bar H}^i L_i  \rightarrow v~\nu_L/\sqrt{2}+...$.
Thus, the r.h.s. of Eq. (\ref {neutrino}) 
can indeed be thought of as a ``weak colorless'' state of 
confining  $SU(2)_L$; moreover, this  state 
corresponds to the left-handed neutrino  of the Standard Model.

Once the bubble is formed, the ``weak colorless'' state $ {\bar H}^i L_i$
can appear in the confining phase inside of the bubble. 
This state, as we established above, carries 
leptonic  charge. 
There is nothing that keeps this  ``weak colorless state''  
within the hot Brane Universe.  Thus, it will be able to 
escape out into the  higher-dimensional space.  
This process would seem as a leptonic  charge non-conserving 
phenomenon to  a four-dimensional observer living in the Brane Universe
(see Fig. 3).   
The same applies to all the other standard model particles. Each of them
can be thought of as ``weak colorless'' bound states \cite {tHooft}. 
Some of them  are listed below:
\begin{eqnarray}
e_L {\rm ~in~ Higgs~phase}~<=>~\varepsilon_{ij}H^iL^j
~{\rm in~ Confinement
~phase}~;  \nonumber \\
u_L  {\rm ~in~ Higgs~phase}~<=>~~~~
{\bar H}^i Q_i ~{\rm in~ Confinement
~phase}~;              \nonumber \\
d_L   {\rm ~in~ Higgs~phase}~<=>~                 
\varepsilon_{ij}H^iQ^j~{\rm in~ Confinement
~phase}~;         \nonumber \\
Z^0  {\rm ~in~ Higgs~phase}~<=>~~                   
{\bar H}D_{\mu} H  ~{\rm in~ Confinement
~phase}~.  \nonumber \\
\label{states}
\end{eqnarray}
\begin{figure}[h]
\epsfbox{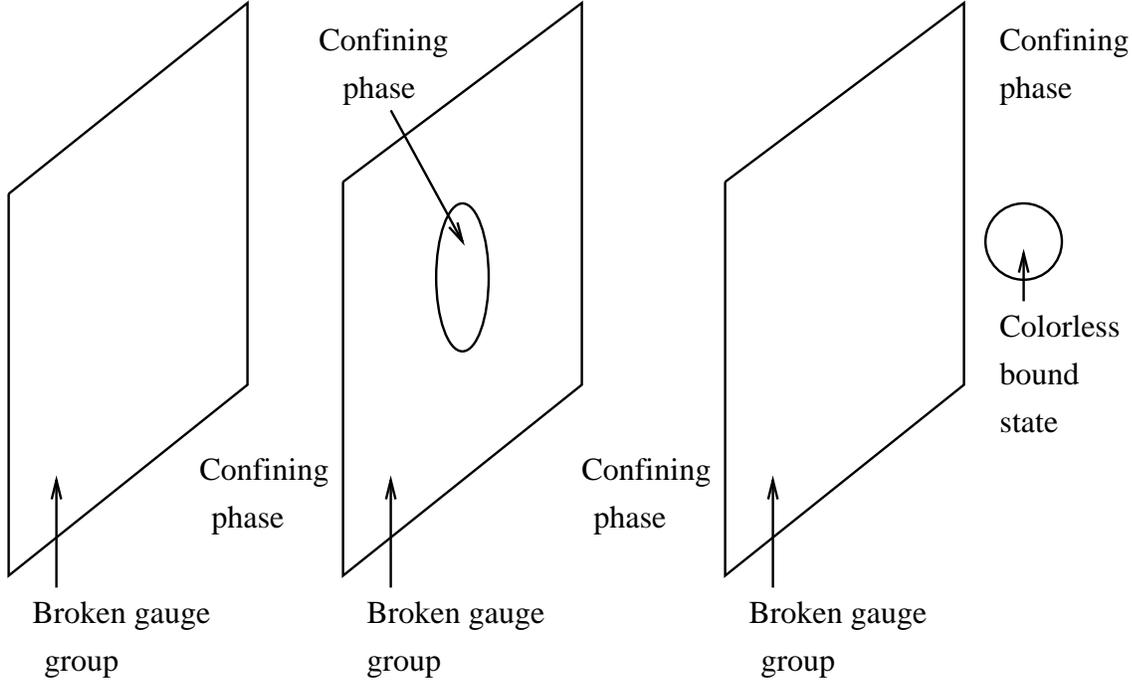}
\vspace{0.3in} 
\caption{{\small Production of confining bubbles}} 
\label{fig3}
\end{figure}   
Here, $Q$ denotes the left-handed up and down quark dublet.
Some combinations of these states,  such as 
(here we suppress all the Lorentz indexes and gamma matrices)
\begin{eqnarray}
\varepsilon_{abc}~{\bar H}^i Q^a_i d^b_R d^c_R, ~~~\varepsilon_{abc}
u_R^a d_R^b d_R^c, 
\label{combinations}
\end{eqnarray}
will be created as ``Standard Model colorless''   
excitations inside of those bubbles
and, as a result, they will escape
our brane at high enough temperatures or energies. 
Evidently, they will be able to carry 
the corresponding global charges, such as lepton or baryon number,
away from the mother brane. 
This will make a four-dimensional observer think that
the global quantum numbers are not  
conserved at high temperatures or energies in the Brane Universe. 
In the next section we
discuss how these processes might  lead to the baryon asymmetry in
the Brane Universe.
\vspace{0.2in} \\
{\bf 3. Baryon Asymmetry in  the Brane Universe} 
\vspace{0.2in}

In this section we argue that the baryon number non-conservation 
mechanisms discussed above might lead to a new  
approach to baryogenesis in the Brane Universe.
We discuss two possible mechanisms. 
The first one is based on the fact that 
C and CP asymmetric branes can treat 
baryons and antibaryons differently.
As a result, the rate to capture a baryon on a baby brane differs from
that for an antibaryon. Thus, net baryon charge accumulation is possible
if the system is out of equilibrium.
This scenario is discussed in subsection 3.1.   
The second scenario is based on production of the baryon number excess 
in a collision of two different Brane Universes after inflation.
This scenario  emerges naturally within  the
recently-proposed ``Brane Inflation'' 
framework \cite {DvaliTye}. The corresponding
discussions are given in subsection 3.2\footnote{Some  more conventional
scenarios of baryogenesis in theories with low $M_{P}$ were discussed in
\cite {BD}. These,  however, are not related to  the present work.}.
\vspace{0.1in} \\
{\it 3.1.  Baryon Asymmetry on the Asymmetric Brane} 
\vspace{0.1in}

As we discussed above, the baby branes and/or confining bubbles
will carry some baryonic  charge off our brane. The very same 
processes will be happening with antibaryons which  
will be taken away from the brane by the same mechanism.
If the theory at hand does not distinguish between 
baryons and antibaryons, then the net charge carried 
away from our brane will average to zero. However,
there is a possibility that the brane actually do 
distinguish between baryons and antibaryons if C and CP are broken. 
In particular, if the rate to capture a baryon on 
a baby brane  differs from the corresponding rate for antibaryons, 
then the accumulation of the  net baryonic  charge 
on our brane will be possible in non-equilibrium processes
\cite {Sakharov}.  

Let us consider a toy model which demonstrates 
how this asymmetry can arise. 
Consider a scalar field $\chi$ which forms a  
four-dimensional ``brane'' embedded 
in five-dimensional space-time. 
Let us say the profile of this soliton  is given by
the familiar ``kink'' solution:
\begin{eqnarray}
\chi= v~ {\rm tanh} (m x_5),
\label{profile}
\end{eqnarray}
where $m^{-1}$ defines  the thickness of the brane
and $v$ stands for the VEV of the corresponding quantum field. 
Consider two five-dimensional fermions coupled to $\chi$:
\begin{eqnarray}
{\cal L}_{\rm int} = \chi~( g_1 {\bar \psi}_1 \psi_1 + g_2 
{\bar \psi}_2 \psi_2)+m_0~\psi^{\rm T}_1 C^{(5)}\psi_2~+~{\rm 
other~~terms}~,   
\label{fivedim}
\end{eqnarray}
where $m_0$ stands for some mass parameter and $ C^{(5)}$ denotes
the charge conjugation matrix in five-dimensional space-time,
$ C^{(5)}\equiv C~\gamma_5$. 
This theory has the  symmetry:
$\psi_1\rightarrow {\rm exp} (i\alpha) \psi_1$,
$\psi_2\rightarrow {\rm exp} (-i\alpha) \psi_2$. We identify this symmetry
group with $U(1)_{\rm baryon}$, thus  
$\psi_1$ and  $\psi_2$ carry opposite baryonic  charges. 
It is well known that each of these fermions give rise  
(in the massless limit) to a single
chiral zero-modes localized on the brane:
\begin{eqnarray}
{\psi}(x)\equiv  {\psi}^0_1(x) ~{\rm exp} \Big ( -\int_0^{x_5} 
g_1 \chi(z)dz \Big ),
~~~~~
{\psi}_c (x) \equiv  {\psi}^0_2(x) ~{\rm exp} \Big ( -\int_0^{x_5} 
g_2 \chi(z)dz \Big ).  
\label{zeromodes}
\end{eqnarray}  
From the point of view of the brane worldvolume field theory 
these chiral fermions can be identified with the worldvolume baryon 
${\psi}$ and  antibaryon   ${\psi}_c $ (in Weyl notations)\footnote{
Switching on small mass $m_0<<g_{1,2}~v$ does not change 
the qualitative picture.}.
In the low-energy theory the ``charge conjugation''
symmetry ${\psi} \rightarrow {\psi}_c$  is broken since $g_1\neq g_2$.
This results in difference between the localization widths
for ${\psi}$ and  ${\psi}_c$ which are given by
$\Delta \propto 1/g_1$ and ${\bar \Delta}\propto 1/g_2$ respectively. 
For instance, the width for (left handed) baryon can be 
made smaller than that for antibaryon ($g_1 > g_2$).  
Then, at energies $\Delta^{-1}< E< {\bar \Delta }^{-1}$ 
the antibaryon ${\psi}_c$
can be ``stripped off'' the brane, while 
the baryon   ${\psi}$ would still be localized.
This toy example explicitly shows how the brane can be ``C-asymmetric''.
For generating net baryon charge, however, CP breaking is also
required. Assuming  that this is the case, (i.e. there are 
some explicitly CP-non-invariant terms in (\ref {fivedim})), 
we expect that the probability for baryons to be captured by a 
baby brane is different
than that for antibaryons (though, this process  
is more difficult to quantify).
As a result, the baby branes 
will be able to remove from our world more antibaryons than baryons. 
Thus, the worldvolume observer will eventually see the net baryon asymmetry
provided that ``evaporation'' into the baby branes is an out-of-equilibrium
process. Such a out-of-equilibrium condition  may emerge  
for instance from the
reheating due to collisions  of two Brane Universes.
 
Note  that in this toy model there are bulk states which 
carry baryon number. They  are Kaluza-Klein states of the original
fermions $\psi_1$ and $\psi_2$. These states can mediate baryon number 
exchange between different branes. However, they are heavy,
and the corresponding interactions are exponentially suppressed
by the brane separation (see related discussions in 
\cite {SavasNima,NimaSh})\footnote{In particular, 
based on the similar effect, N. Arkani-Hamed and M. Schmaltz 
\cite {NimaSh} proposed the mechanism for the proton decay suppression.}. 
Moreover, in realistic models  due to the bulk confinement
(which we have ignored in this toy example)  these 
heavy states can only propagate within the 
bulk ``colorless  Hadrons''. 

Finally,  we would like to discuss the issue of the overclosure
of the Universe by baby branes in such a scenario\footnote{We are
grateful to V. Rubakov for rising this issue.}. In order to generate the
net baryon asymmetry on our brane, not all the baby branes should
return to it. If they stay in the bulk, they will look as a sort 
of dark matter with TeV mass. 
If we assume roughly one unit of baryon number captured per
baby brane, their number density would be so large that they would
overclose the Universe. 
However, there are several ways to  avoid this problem.
The most straightforward is to 
notice  that the 
baby branes need not stay in the bulk,  but
rather can be ``discharged'' on some other distant brane (like ours, or
even larger dimensionality).
In such a case the energy density of baby branes will be converted into the 
distant brane tension and will  be absorbed into the  effective over-all
cosmological term
\begin{equation}
\Lambda_{\rm eff} = \sum_i \sigma_i + \Lambda_{\rm bulk}V_N,
\end{equation}
where $\Lambda_{\rm bulk}$ is the bulk cosmological constant and the
summation is over all branes. The probability that the baby brane
encounters a bigger brane and gets discharged there needs further 
quantification within more  realistic models. 
\vspace{0.1in}  \\
{\it 3.2.  Brane Inflation and Baryon Asymmetry} 
\vspace{0.1in}

In this subsection we discuss the  mechanism of baryogenesis
which naturally arises  within the Brane Inflation framework 
\cite {DvaliTye}. According to the general Brane Universe scenario, 
we live on a brane or a set of overlapping branes. 
The later possibility is supported by
D-brane constructions in which the
existence of a non-Abelian gauge group requires 
a number of parallel D-branes sitting  on top of each other
\footnote{Cosmological scenarios within Horava-Witten type theories 
\cite {HW} were recently discussed in Ref. \cite {Ovrut}.}.

Before supersymmetry is broken branes are  BPS
states with zero net force between them. This is certainly true
for two (or more) parallel D-branes, where
the gravitational and dilaton attraction is exactly canceling with 
the repulsion mediated by Ramond-Ramond fields \cite {Polchinski}. 
Similar examples can be constructed for
field-theoretic branes, topological solitons 
\cite {shifman}\footnote{For a reviews on 
supergravity solitons 
see, e.g., \cite {Cvetic}, \cite {Duff}.}.   
However, in the real world supersymmetry must be broken and dilaton
should  be stabilized. 
Thus,  we expect a non-zero net force
between branes. 
The general expression for a potential between two such 
parallel branes embedded in $N>2$ transverse dimensions 
at large distances ($r>>M^{-1}$) takes the following form: 
\begin{equation}
V(r) = M^4\left (d+ {b_je^{-rm_j} - 1 \over (M~r)^{N - 2}} \right ).
\label{potential}
\end{equation}
The constant term $d$ comes from the short-range brane-brane  
interaction. In fact, it accounts for  interactions  between particles
localized on different branes, whose wave-functions only can overlap if
branes intersect. The potential is normalized as 
$V(\infty) = 2\sigma$, $\sigma$ being the brane tension. 
Yukawa potentials in (\ref {potential})  
come from the exchange of 
heavy bulk modes with masses $m_j$, and the power law interaction
comes from the bulk gravitational attraction. 
If the D-brane picture is adopted, then
$m_j$'s should be understood 
as masses of dilaton and Ramond-Ramond fields. 
Regardless of what is the actual realization of 
branes, be it the D-brane picture or field theory soliton context, 
the potential in  (\ref {potential}) describes adequately 
interactions between those  objects. 
The model-dependent quantities  are parameterized by coefficients
$d, b_j$ and $m_j$.  
These parameters determine the minimal separation 
$r_{\rm vac}$ at which the branes are stabilized in 
the lowest-energy state.
If $r_{\rm vac} < M^{-1}$,   
the separation between the branes is smaller than the typical 
size at which the branes could fluctuate. Thus, the branes
effectively sit on top of each other. 
As a result,  the particles
localized on these two branes are effectively 
shared by both of them. Below we will concentrate on the following 
alternative possibility. Let us assume that  $r_{\rm vac} >> M^{-1}$. 
In this case particles localized on two different branes 
have no overlap. Thus, they 
belong to either of branes,  but are not shared among 
them. These two worlds can 
communicate to each other by exchanging 
bulk fields. If  these interaction
preserve global charges, $B$ and  $L$ charges are conserved
separately on each branes.

Let us see how this picture is affected by 
the dynamics of the brane inflation \cite {DvaliTye}. 
Once the  branes are separated by a distance 
$r>> r_{\rm vac}$, the nonzero potential energy between the branes 
gives rise to the four-dimensional effective cosmological constant that
drives inflation \cite {DvaliTye}. This constant can be defined as follows:
\begin{equation}
\Lambda_{\rm eff} = V(r) + \Lambda_{\rm bulk}V_N,
\label{const}
\end{equation}
where $\Lambda_{\rm bulk}$ is the bulk cosmological constant and $V_N$
is the volume of the extra compactified space. 
Nearly zero value  of the cosmological constant that is observed today
implies that 
\begin{equation}
\Lambda_{\rm vac} = V(r_{\rm vac}) + \Lambda_{\rm bulk}V_N \simeq 0. 
\end{equation}
Thus, according to Eqs. (\ref {potential},\ref {const}), 
the  four-dimensional vacuum
cosmological constant  will be nonzero for any $r \neq r_{\rm vac}$. 
This potential energy will drive inflation, 
the exponential growth of the three non-compact dimensions\footnote[2]{
The size of the extra dimensions will not be affected by this growth
provided that the mass of the radius modulus is at least ${\rm mm}^{-1}$
\cite {DvaliTye}. 
Some possibilities of primordial KK inflation with
change of $R$ were studied in \cite {Cosmology}.}. 
The next crucial thing is to note that 
for $r >> r_{\rm vac}$ the potential  (\ref {potential}) 
is a very flat function of $r$. As a result, 
the branes fall very slowly on each other. Thus, during this process the
Universe is dominated by the potential energy which in fact 
triggers inflation in non-compact dimensions. 
We should also emphasis  that the compact dimensions will not
inflate since the effective Hubble size is never smaller than the
size of the compact dimensions \cite{DvaliTye}. 
From the point of view of
an effective four-dimensional 
theory this process is equivalent to slow rolling of 
a scalar field, an inflaton
\begin{equation}
\Phi = rM^2 .
\end{equation}
This field, according to (\ref {potential}) has
a very flat potential. The quantity $\langle \Phi\rangle  =
r_{\rm vac}M^2 $ is just the vacuum expectation 
value of the inflaton today. 

The end of inflation is determined 
by the value of $\Phi$ which 
breaks either  of the standard slow-roll conditions 
$V'M_P/V < 1, V''M_P^2/V < 1$
(see \cite{DvaliTye} for details). 
The epoch in which we are interested in 
starts right at this point of the evolution. 
We will argue below that after the branes collide
and reheat each other, 
the net baryonic  charge  can be induced on our brane.

One possible scenario emerges when  
the branes get stabilized after the collision at some
large distance $r_{\rm vac} >> M^{-1}$. 
This is going to be the case if the  branes repeal at 
short distances. For instance, this condition 
can  be realized  
within the D-brane construction 
if dilaton becomes heavier than the corresponding Ramond-Ramond
field $m_D \sim M >> m_{\rm RR}$. 
As a consequence, when $r << m_{\rm RR}^{-1}$ the Ramond-Ramond
repulsion takes  over and branes get stabilized at 
$r_{\rm vac} \sim m_{\rm RR}^{-1}$\footnote[1]{The 
value of $m_{\rm RR}^{-1}$ can control the 
reheating temperature after ``Brane Inflation'' \cite {NimaSavasGia},
this  clarifies the issue of reheating temperature 
discussed by T. Banks, M. Dine and A. Nelson (see the reference in
\cite {jgro}).}.

Let us follow this scenario more closer. 
The potential energy of the Universe
during inflation can be estimated as follows:
\begin{equation}
\Lambda_{\rm eff} (r >> r_{\rm vac}) \sim M^4 
\Big ({m_{\rm RR} \over M}\Big )^{2-N}.
\end{equation}
This amount of energy will 
transform into the energy of colliding branes
after inflation. Let the wave-function of a set of particles 
localized on
our brane be $\psi_{\rm our}(x_{\mu})$, likewise, the wave function 
of a set of some different particles living on the other brane be 
$\psi_{\rm other}(y_{\mu})$. There is a $U(1)_{\rm our}$ 
baryon number symmetry on our brane
\begin{equation}
\psi_{\rm our}(x_{\mu}) \rightarrow {\rm e}^{iQ_{\rm our}\alpha}
\psi_{\rm our}(x_{\mu}).
\end{equation}
Likewise, there is a similar 
$U(1)_{\rm other}$ symmetry  on the
other brane
\begin{equation}
\psi_{\rm other}(x_{\mu}) \rightarrow {\rm e}^{iQ_{\rm other}\beta}
\psi_{\rm other}(x_{\mu}). 
\end{equation}
When branes are separated, these are two {\it different}
symmetries.  In the effective
four-dimensional theory, this simply means that the interactions
that break $U(1)_{\rm our}\otimes U(1)_{\rm other}$ are suppressed
as follows:
\begin{equation}
(\psi_{\rm our}^*)^{Q_{\rm
other}}~ e^{-rM}~  (\psi_{\rm other})^{Q_{\rm our}}.
\label{overlap}
\end{equation}
However, once the  branes come on top of each other,
the suppression goes away. As a result, we are left 
with the only one conserved charge $Q = Q_{\rm other} +
Q_{\rm our}$. 

During inflation particles are  inflated away on both
branes and the expectation values of the operators $Q_{\rm other}$ and 
$Q_{\rm our}$ vanish.
When the branes collide part of their energy is spent on  
creation of particles, baby branes and/or bubbles (see Fig. 4). 
Since the total charge $Q$ is conserved, the net 
charge produced on the 
both
branes should be  zero.  However, 
during the  non-equilibrium collision process the 
branes overlap. Thus, $Q_{\rm other}$ and $ Q_{\rm our}$ will 
{\it not} be separately conserved, and 
it might happen that in some reactions 
$ \Delta Q_{\rm our} = - \Delta Q_{\rm other}
\neq 0$. Thus, the net global charges will be left
on each branes. 
In addition to this effect, 
some charge will be carried away by the baby
branes and/or the confining bubbles as discussed in the previous
sections.
We can briefly summarize the process described above as follows:
The  branes, while colliding,  spend a very little
time on top of each other. After that, they just 
``bounce back'' and start to oscillate
about the equilibrium point $r_{\rm vac}$. 
If C and CP symmetries  are broken during the 
brane collisions,
the couplings (\ref {overlap}) allow ``charges''
to be ``exchanged''  among $\psi_{\rm our}$ and $\psi_{\rm other}$
during the short time moment of the collision.
Thus, it might happen that one charge is produced 
in inflaton decays  in excess and the other one in deficit (see
the example below).  Once the collision happened, these couplings 
switch-off almost instantly, and as a result,  
the values of nonzero charge asymmetries  
$ \Delta Q_{\rm our} = - \Delta Q_{\rm other} \neq 0$ freeze-out. 
This, in particular, happens since the couplings  (\ref{overlap})
vanish almost instantly  and the charges become separately
conserved on two different branes.

The qualitative discussions given above can  be made more precise
by considering a simplified toy model.
Consider two types of fermions, let us call them  $B_j$ and
$D_A$. $B_j$'s are  localized on our brane and carry baryon number
($U(1)_B$). 
$D_A$'s, on the other hand,  are localized on 
a distant brane and carry the corresponding global  charge ($U(1)_D$). 
Given the exponential suppression of the overlap of their wave-functions,
a part of the  effective four-dimensional Lagrangian 
for these fermions can be written as follows: 
\begin{eqnarray}
{\cal L}_{\rm int} = c_{ij}(\Phi)~ B_i^c B_j~ +~
c_{AB}(\Phi)~ D^c_A D_B~ 
+~\lambda_{iA}(\phi)~ e^{-{\Phi\over M}}~ D_A^cB_i~+~
{\bar \lambda}_{iA}~e^{-{\Phi\over M}}~ B^c_i D_{A} \nonumber \\ 
+ c_{ijkm}(\Phi)~ B_i^c B_jB_k^c B_m~ + 
c_{ABCD}(\Phi)~ D^c_A D_B D^c_C D_D  + {\rm other
~interactions~ +~ h. c.}.
\label{BD}
\end{eqnarray}
Here,  $\Phi\equiv  M^2r$ denotes a brane-separation modulus field,
the inflaton, and $c$'s and $\lambda$'s are some polynomial 
functions of $\Phi$  in which  C and CP  violations are 
encoded. $B^c_i$ and $D^c_A$ stand for charge conjugated 
fields. 
Note that interactions of $\Phi$ with the fields on the same
brane need not be exponentially suppressed. In some cases, these 
interactions arise after  integrating out the bulk modes
(e.g. open string modes stretched between two branes) which acquire masses
due to the VEV of $\Phi$  and have direct couplings to  the light 
modes on each brane. 
On the other hand,  all the overlapping terms which  break
$U(1)_B\otimes U(1)_D$ symmetry explicitly must be exponentially
suppressed (since,  by the assumption, there are no light bulk modes
with these charges).

Thus, when branes are well separated $\Phi >> M$, the overlap
terms are suppressed and  
the  Lagrangian has two independent $U(1)$-symmetries. 
One of them  acts on $B$'s  and can be regarded  as baryon 
number symmetry in our brane. When branes come closer, however, 
the overlap terms do not vanish. As a result,  
we are left with one common fermion-number Abelian symmetry group
$U(1)_F$. This last  conserves the ``total charge'' of the 
branes $ Q\equiv Q_{\rm B}+Q_{\rm D}$. 
Let us now turn to the particles which 
are being created in the inflaton decay. 
This decay, as we just mentioned,  conserves the total charge $Q$.
However, the individual charges, $Q_{\rm B}$ and $Q_{\rm D}$
are not conserved. 
Therefore, the rate for baryon number creation (e.g. in
two-body decays) 
\begin{equation}
\Phi \rightarrow B_i + D_A^c, ~~~~~~ \Phi \rightarrow B_{i}^{c*} + D_A^*~,
\end{equation}
and, likewise, the rate for  antibaryon number creation, are different.  
Thus, although $Q$ is conserved in the inflaton decays, individually
$Q_B$ and $Q_D$ will not be conserved if both C and CP  are broken. 

Note that there might  exist  an additional
source of physical CP  violation due to ``time interface'' which 
can arise 
as a result of  the time-dependent VEV of $\Phi$
and  different dimensional operators present in the $c$ and
$\lambda$ functions.
These contributions are clearly very model-dependent and we will
not attempt  to quantify them here. An important outcome, however, is
that in general,  the rate to produce baryons in the inflaton decays
differs  from the same rate for antibaryons if C and CP are broken.
Therefore, the  nonzero value of
$\Delta Q_{\rm B} = - \Delta Q_{\rm D}$ will be produced. 
When the branes bounce back after the collision, the inflaton
VEV sharply increases.  Thus, the $U(1)_B$  violating terms in (\ref {BD})
switch  off and  the baryon generation process stops before the 
system equilibrates. As a result, the accumulated net baryonic  
charge $\Delta Q_{\rm B}$  freezes-out. 
Thus, our brane  will be carrying
the  net baryon  number  after the system comes to  equilibrium.
\vspace{0.2in} \\
{\bf 4. Discussions and Conclusions}
\vspace{0.2in}

Summarizing, we have shown  that
the processes with non-conservation of global charges,
such as baryon  and lepton number,  will be taking place 
in the  Brane Universe. A number of different scenarios of 
non-conservation of the global charges were proposed.
The basic idea is that the global charges can be carried 
away from the Brane Universe by baby branes and/or confining
bubbles. These mechanisms  
do not depend on a particular
model of interactions realized on the brane. Thus,  
they are model-independent and inherent for
the Brane Universe scenario. 

We have also presented two independent scenarios 
of baryogenesis in the Brane Universe.
These approaches are based  upon 
properties of higher-dimensional space-time, 
as well as on non-equilibrium nature of evolutionary processes 
taking place  in the  Brane Universe. Highly
non-equilibrium and violent nature of these processes, be it scatterings
of various branes of  different dimensionality or
strong quantum fluctuations of hot branes, makes it 
complicated to perform precise analytical  studies. Hopefully, these
mechanisms  can be tested  quantitatively in
numerical simulations of various toy and/or realistic models.
\vspace{0.2in} \\
{\bf Acknowledgments}
\vspace{0.1in}

The authors are grateful to Nima Arkani-Hamed, Savas Dimopoulos,  
Glennys  Farrar and especially Valery Rubakov for valuable  discussions
and comments.  
The work of GG was supported in part by the grant NSF PHY-94-23002. 
\vspace{0.5in} \\
\begin{figure}[h]
\epsfbox{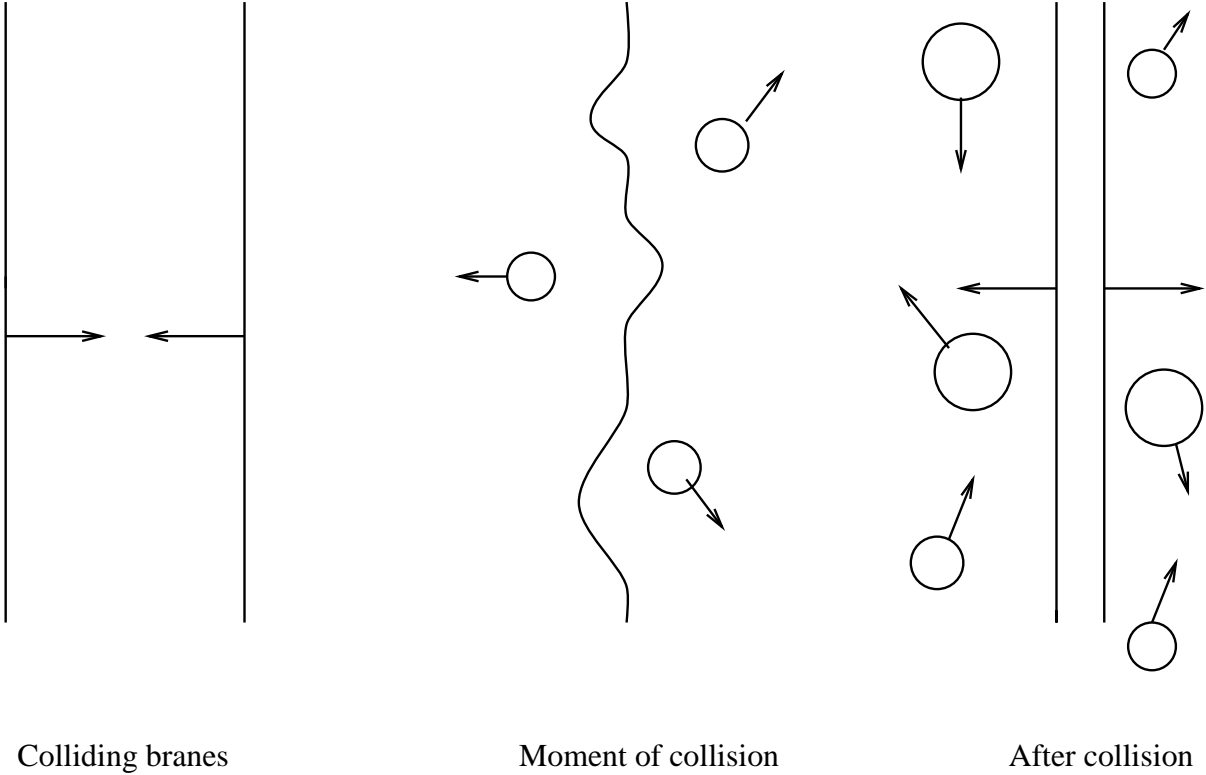}
\vspace{0.3in} 
\caption{{\small Colliding branes produce baby branes}} 
\label{fig4}
\end{figure}

\end{document}